# Effects of extrinsic point defects in phosphorene:

# B, C, N, O and F Adatoms

Gaoxue Wang[1], Ravindra Pandey[1]*, and Shashi P. Karna[2]

[1]Department of Physics, Michigan Technological University, Houghton, Michigan 49931, USA

[2]US Army Research Laboratory, Weapons and Materials Research Directorate, ATTN: RDRL-WM, Aberdeen Proving Ground, MD 21005-5069, USA

(April 15, 2015)

Email:   gaoxuew@mtu.edu
         pandey@mtu.edu
         shashi.p.karna.civ@mail.mil

**Abstract**

Phosphorene is emerging as a promising 2D semiconducting material with a direct band gap and high carrier mobility. In this paper, we examine the role of the extrinsic point defects including surface adatoms in modifying the electronic properties of phosphorene using density functional theory. The surface adatoms considered are B, C, N, O and F with a [He] core electronic configuration. Our calculations show that B and C, with electronegativity close to P, prefer to break the $sp^3$ bonds of phosphorene, and reside at the interstitial sites in the 2D lattice by forming $sp^2$ like bonds with the native atoms. On the other hand, N, O and F, which are more electronegative than P, prefer the surface sites by attracting the lone pairs of phosphorene. B, N and F adsorption will also introduce local magnetic moment to the lattice. Moreover, B, C, N and F adatoms will modify the band gap of phosphorene yielding metallic transverse tunneling characters. Oxygen does not modify the band gap of phosphorene, and a diode like tunneling behavior is observed. Our results therefore offer a possible route to tailor the electronic and magnetic properties of phosphorene by the adatom functionalization, and provide the physical insights of the environmental sensitivity of phosphorene, which will be helpful to experimentalists in evaluating the performance and aging effects of phosphorene-based electronic devices.

Phosphorene is the monolayer form of the black phosphorus. Since the interlayer interaction in the bulk black phosphorus is dominated by the van der Waals forces, phosphorene could be obtained by exfoliating from the bulk lattice [1-4]. In fact, a few layer of phosphorene is successfully exfoliated and exploited for applications in electronic devices [1]. Recently, it has been demonstrated that phosphorene-based transistors possess a larger current on/off ratio compared to graphene-based transistors and higher charge mobility than $MoS_2$-based devices [1, 5]. Furthermore, tunable band gap [6, 7], directional dependent conductance [8], and fast photo-response have been predicted for phosphorene [9, 10], thereby inducing interest amongst scientists for its novel applications in devices at nanoscale [11, 12]. It should be noticed that encapsulation of phosphorene by BN monolayers has demonstrated to improve stability of phosphorene in air for device applications [13].

It is well-known that the exfoliation or growth processes can introduce defects and impurities in two-dimensional (2D) materials which can dramatically alter the electronic, thermal and mechanical properties of the pristine counterparts. Vice versa, a deliberate introduction of defects can be a possible approach to modify the properties of the pristine materials. For example, ion or electron irradiation can introduce intrinsic point defects, e.g. vacancies and Stone-Wales (SW) defect in graphene [14, 15], which could enrich its properties to act as a building block for devices with new functionalities [16, 17]. Besides intrinsic defects, extrinsic defects such as adatoms are shown to be important to design graphene based devices with dedicated properties [18]. Also, exposing of 2D $MoS_2$ to ultraviolet ozone is found to functionalize the monolayer providing nucleation sites for the growth of atomic layers of $Al_2O_3$ [19].

Considering that the scientific work on investigating the properties of phosphorene has recently started, the role of extrinsic point-defects including surface adatoms is still undefined. The adsorption of several adatoms was considered recently [20-24], but the underlying mechanism of the different behaviors of the adatom on phosphorene is not mentioned. In this paper, we focus on the adsorption of a series of adatoms from B, C, N, O, and F which provide an interesting variation in the number of valence electrons with a [He] core and the electronegativity. B is $s^2p^1$, C is $s^2p^2$, N is $s^2p^3$, O is $s^2p^4$ and F is $s^2p^5$. The values of electronegativity for B is 2.0, C is 2.5, N is 3.1, O is 3.5 and F is 4.1 [25]. The electronegativity of P is 2.1 [25]. We will calculate their geometric structure and electronic properties and will compare the results with those on graphene and silicene to gain insight in the adsorption mechanism. We will show that the adsorption behavirour of B and C is totoally different from that of N, O and F adatoms, and the adsorption mechanism is dorminated by the electronagetivity of the adatom and the surface electronic structure of phosphorene. In Sec 2, we give a brief description of the computational model. Results and discussion are given in Sec. 3. Finally, a summary is given in Sec 4.

The electronic structure calculations were performed using the norm-conserving Troullier-Martins pseudopotential implemented in the SIESTA program package [26]. The Perdew-Burke-Ernzerhof (PBE) [27] exchange-correlation functional to density functional theory was employed, which has been shown to correctly describe the adsorption of adatom on graphene [28], silicene [29], and the adsorption of O adatom on phosphorene [22]. The energy convergence is set to $10^{-5}$ eV, and the residual force on each atom is smaller than 0.01 eV/Å during structural optimization. The mesh cutoff energy was chosen to be 500 Ry. A double-$\zeta$ basis including polarization orbitals was used.

The (4×5) supercell with 80 atoms was used to simulate the pristine phosphorene. The length of the supercell was (18.6 Å ×16.7 Å), and the vacuum distance normal to the 2D lattice was chosen to be 20 Å to eliminate interaction between the replicas. A single adatom was added in the supercell yielding adatom concentration of ~$3.23\times10^{13}$/cm$^2$. The reciprocal space was sampled by a grid of (4×5×1) *k* points. The tunneling current calculations were based on Bardeen, Tersoff, and Hamann (BTH) formulism [30, 31].

In order to benchmark the modeling elements, the results on the pristine phosphorene were compared with the previously reported results [7, 32, 33]. Figure S1 [34] shows the ground state configuration, band structure and the charge density at the valence band maximun (VBM) and conduction band minimun (CBM) for phosphorene. The bond lengths are calculated to be 2.29 and 2.26 Å, which are consistent with the values of 2.28 and 2.24 Å obtained at the PBE-DFT level of theory [6, 10]. A direct band gap slightly less than 1 eV is predicted for the pristine phosphorene which is in excellent agreemnt with the previous thereoctical reports [7, 32, 33].

We begin with calculations to determine the energy profile of the adatom approaching surface sites of phosphorene (Figure 1). The surface sites considered are (i) hexagonal site (H) - site above the center of hexagonal ring, (ii) top site (T) - above the top phosphorus atom, and (iii) bridge site (B) - above the bridge of the top P-P bond. The energy profile was initially obtained by varying the distance of the adatom to the 'rigid' phosphorene. Later, a full structural optimization was performed to obtain the ground state configuration in which all atoms are allowed to relax.

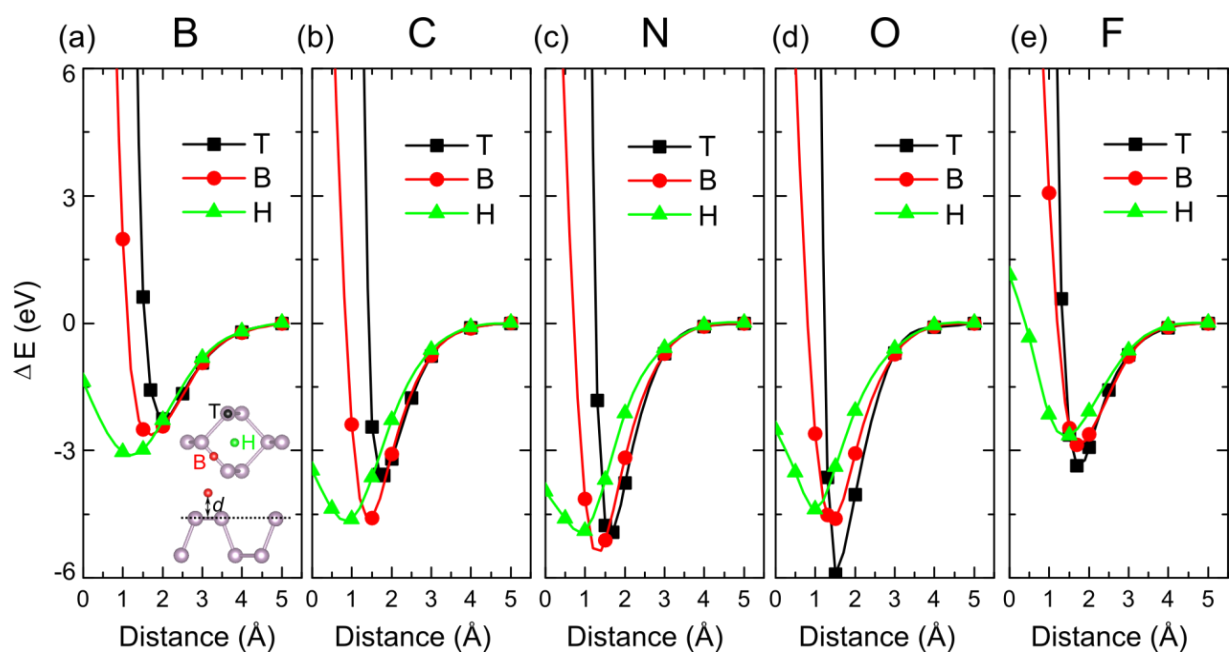

*Figure 1. Energy profile of adatoms approaching the surface of phosphorene at the top (T), bridge (B), and hexagonal site (H): (a) B, (b) C, (c) N, (d) O, and (e) F.*

The calculated ground state configurations of the adsorbed systems are shown in Figure 2. Interestingly, B and C adatoms will break the native P-P bonds and enter the interstitial site of the 2D lattice. The configuration at the top site is found to be 0.75 and 1.27 eV higher in energy for B and C adatom, respectively (Figure S2 [34]), demonstrating that B or C atom prefers to penetrate into the pristine lattice. On the other hand, the interstitial site is found to have higher energy than the surface site for N, O and F atoms (Figure S2 [34]). The results, therefore, show that N, O and F atoms tend to bind the surface P atom without breaking the P-P bonds. The (average) bond lengths of surface adatoms are $R_{P-B}$ (1.96 Å), $R_{P-C}$ (1.80 Å), $R_{P-N}$ (1.68 Å), $R_{P-O}$ (1.54 Å), and $R_{P-F}$ (1.70 Å).

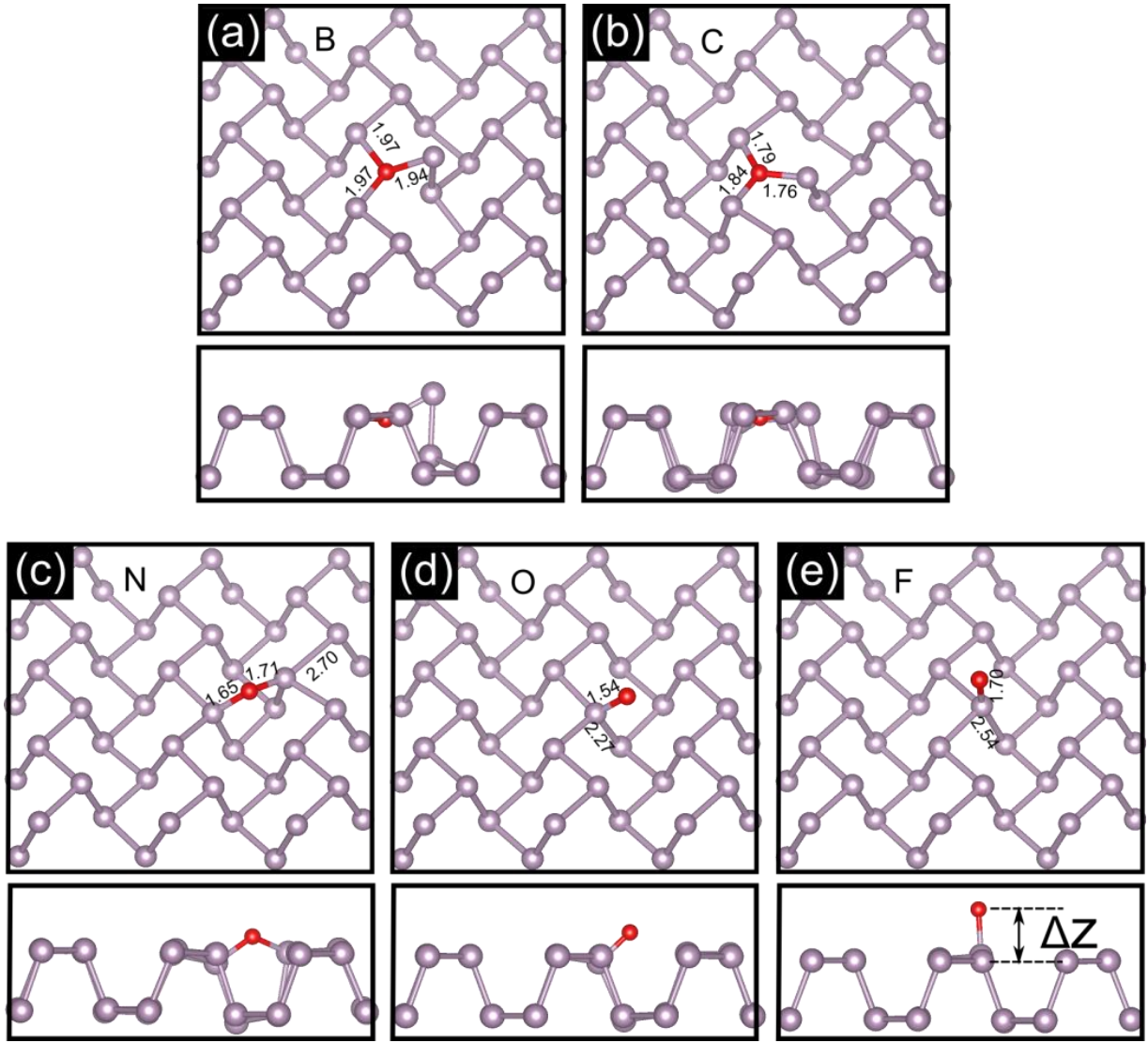

*Figure 2. Adsorption of adatoms with a [He] core on phosphorene: The ground state configurations of (a) B, (b) C, (c) N, (d) O, and (e) F adsorbed phosphorene.*

*Table 1. Adsorption of adatoms with a [He] core on phosphorene: The prefered binding site, distance between adatom to the host P atom (R), binding energy ($E_{binding}$), Voronoi charge ($Q_{Voronoi}$) and magnetic moment ($\mu_B$). Δz is the height of adatom from the surface. Negative Voronoi charge means excess of electron.*

| Adatom | Preferred Site | $R_{host-adatom}$ (Å) | $\Delta z$ (Å) | $E_{binding}$ (eV) | $Q_{Voronoi}$ (e) | Magnetic moment ($\mu_B$) |
|---|---|---|---|---|---|---|
| B ($s^2p^1$) | Interstitial | 1.96 | ≈0.0 | -5.08 | -0.19 | 1.00 |
| C ($s^2p^2$) | Interstitial | 1.80 | ≈0.0 | -5.16 | -0.21 | 0.00 |
| N ($s^2p^3$) | Surface | 1.68 | 0.51 | -2.98 | -0.31 | 1.00 |
| O ($s^2p^4$) | Surface | 1.54 | 0.87 | -4.69 | -0.33 | 0.00 |
| F ($s^2p^5$) | Surface | 1.70 | 1.78 | -2.30 | -0.15 | 1.00 |

B and C atoms reside in the 2D lattice, while N, O, or F atoms stay on the surface. The height of the adatom from the surface (Δz) gradually increases in going from N to F (Table 1). In the equilibrium configurations, some of native P atoms are repelled away by B or C at the interstitial site resulting into lattice distortions. For N, O, or F surface adatoms, the overall lattice structure of phosphorene is maintain. Note that the P-P bond length in the pristine 2D lattice is 2.26 Å. The binding energy of the surface adatom is defined as

$$E_{binding} = E_{total} - (E_{pristine} + E_{atom}) \qquad (1)$$

where $E_{total}$ is the total energy of phosphorene with surface adatom, $E_{pristine}$ is the energy of pristine phosphorene. $E_{atom}$ is the energy of a single adatom and is -97.10 eV, -153.12 eV, -271.94 eV, -440.22 eV, and -665.22 eV for B, C, N, O and F atoms, respectively calculated using the (10 Å×10 Å ×10 Å) unit cell. The adsorption of these light elements is found to be exothermic with a negative binding energy which implies that phosphorene could strongly bind these light elements (Table 1).

In order to understand the site dependency of the light elements with a [He] core, the deformation charge density ($\rho=\rho_{total}-(\rho_{phosphorene}+\rho_{atom})$) of the adsorbed phosphorene is calculated (Figure 3). B and C at the interstitial sites clearly form bonds with three native P atoms. These $sp^2$–like bonds almost lie in a same plane, and the charge density increases in the region between the adatom and P (Figures 3(a), and (b)). On the other hand, N, O and F adatoms appear to form bonds with one or two native P atoms at the surface in their equilibrium configurations. Accumulation of the charge density around the adatom suggests that N, O and F atoms gain electrons from the host P atoms (Figure 3(c), (d) and (e)).

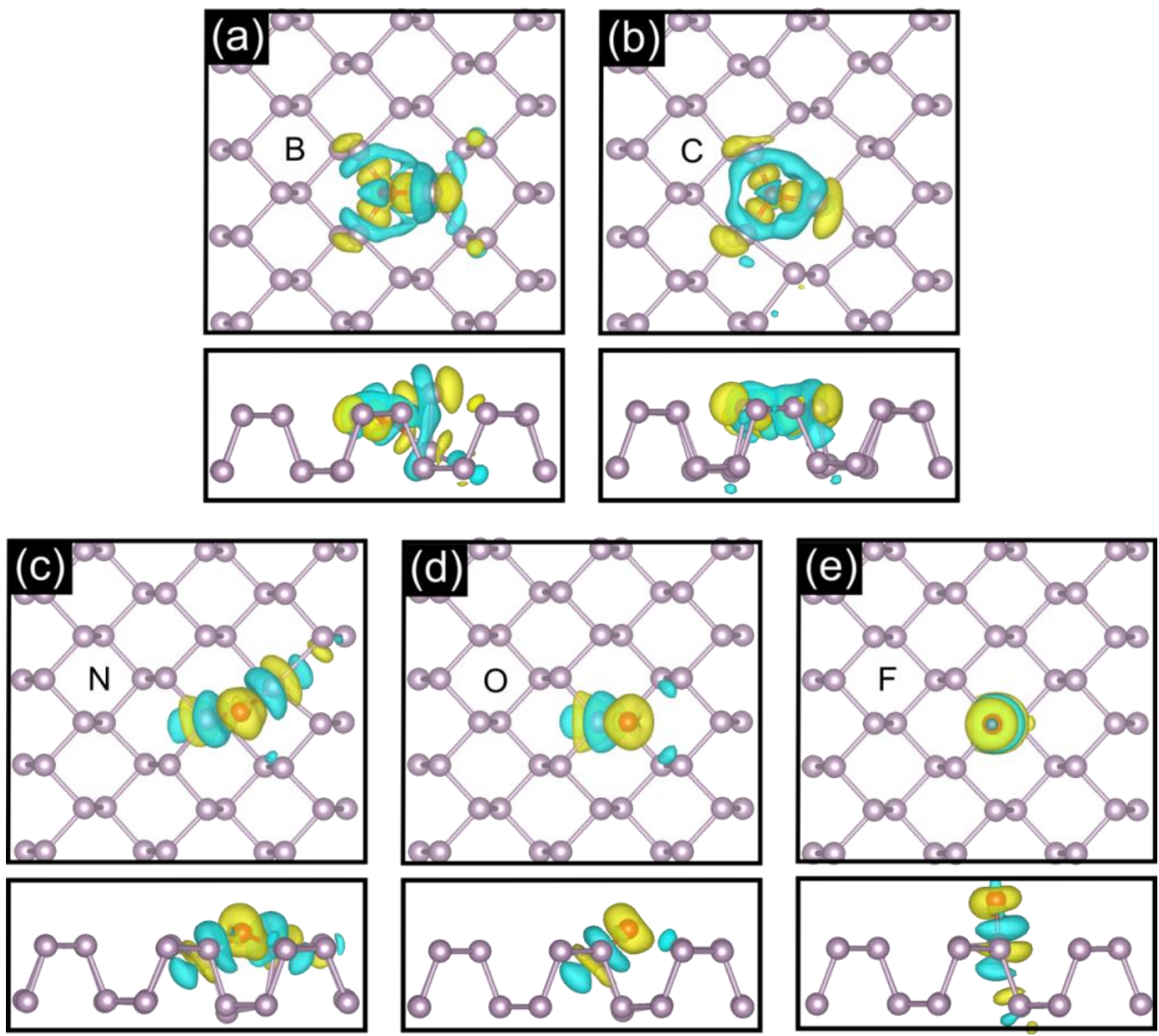

*Figure 3. Deformation charge density of adsorption of adatoms with a [He] core on phosphorene: (a) B, (b) C, (c) N, (d) O, and (e) F. The yellow (blue) region represents accumulation (depletion) of electrons. The isovalue is 0.003 e/Å³.*

A schematic illustration explaining the adsorption of light elements with a [He] core on phosphorene is given in Figure 4. In phosphorene, each P atom (with $s^2p^3$ valence electron configuration) shares three of its valence electrons with the neighboring P atoms forming $sp^3$ bonds while the remaining valence electrons form a lone pair at the surface. Since N, O, and F atoms are more electronegative than the host P atoms, these adatoms tend to attract electrons from the native P atom. For example, F has $s^2p^5$ valence electron configuration and will attract one electron of the lone pair forming bond with one native P atom. O has $s^2p^4$ electron configuration attracting the lone pair of P atom, possibly form P=O bond with the bond length of 1.54 Å [20]. N has $s^2p^3$ configuration and will form bond with two native P atoms. This scenario is clearly reflected in Table 1 showing the calculated bond lengths ($R_{host-adatom}$) in the equilibrium configurations of the adsorbed systems. B and C atoms are close to P atom in terms of the electronegativity, they prefer to form $sp^2$ bonds with the native P atoms by breaking the native $sp^3$ bonds of the pristine phosphorene. We note that the results on oxygen atom on phosphorene are consistent with the results of previous investigations reporting different levels of functionalization of phosphorene by oxygen [22] and stability of phosphorene structures with partial and complete coverage of oxygen [21].

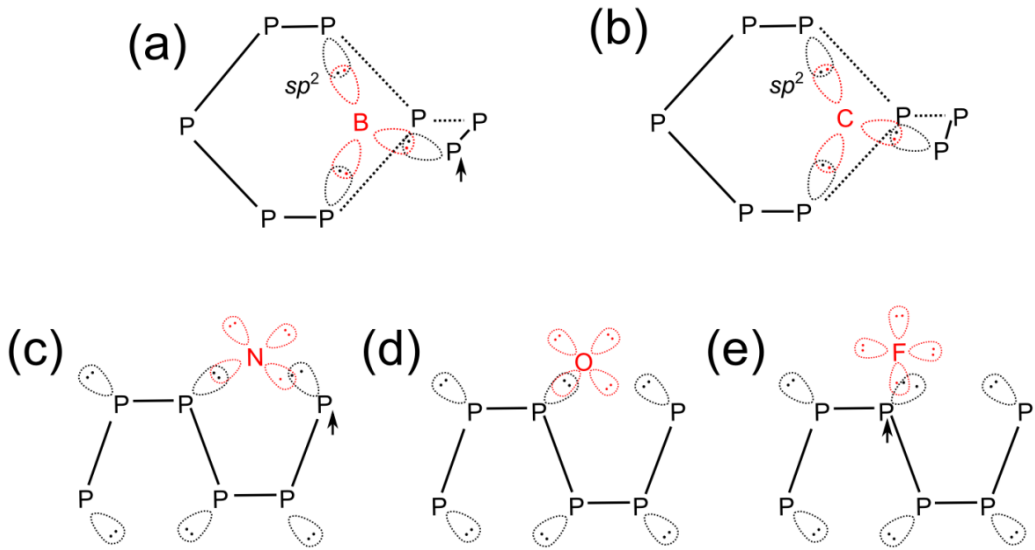

*Figure 4. A schematic illustration of adsorption of adatoms with a [He] core on phosphorene: (a) B, (b) C, (c) N, (d) O and (e) F. For B, N and F, the upaired electron is shown by the arrow.*

Comparing our results on phosphorene with those obtained for graphene [35-37], we find that adatoms prefer surface sites on graphene due to robustness of the $sp^2$ network; B, C, N and O adatoms prefer the bridge site whereas F adatom prefers the top site. The calculated binding energies of B, C, N, O, and F adatoms on graphene are -1.77 [35], -1.4 [36], -0.88 [37], -2.41 [37], -2.01 eV [37], respectively.

It should be noticed that $sp^3$ bonds tend to be more reactive leading to higher binding energies of surface adatoms on phosphorene which is puckered (Table 1). This is also the case for silicene where $sp^3$-like puckered configuration is highly reactive[38]. The calculated results show that B or C adatoms prefer to occupy the lattice sites inducing significant distortion in the 2D lattice of siliciene[38,39]. On the other hand, N, O and F adatoms prefer the surface sites [38-40] including the bridge or top sites without breaking the Si-Si bonds in the 2D lattice. The calculated binding energy of B, C, N, O or F adatoms on silicene is -5.85 [38], -5.88 [39], -5.54 [38], -6.16 [39], and -4.45 eV [40], respectively.

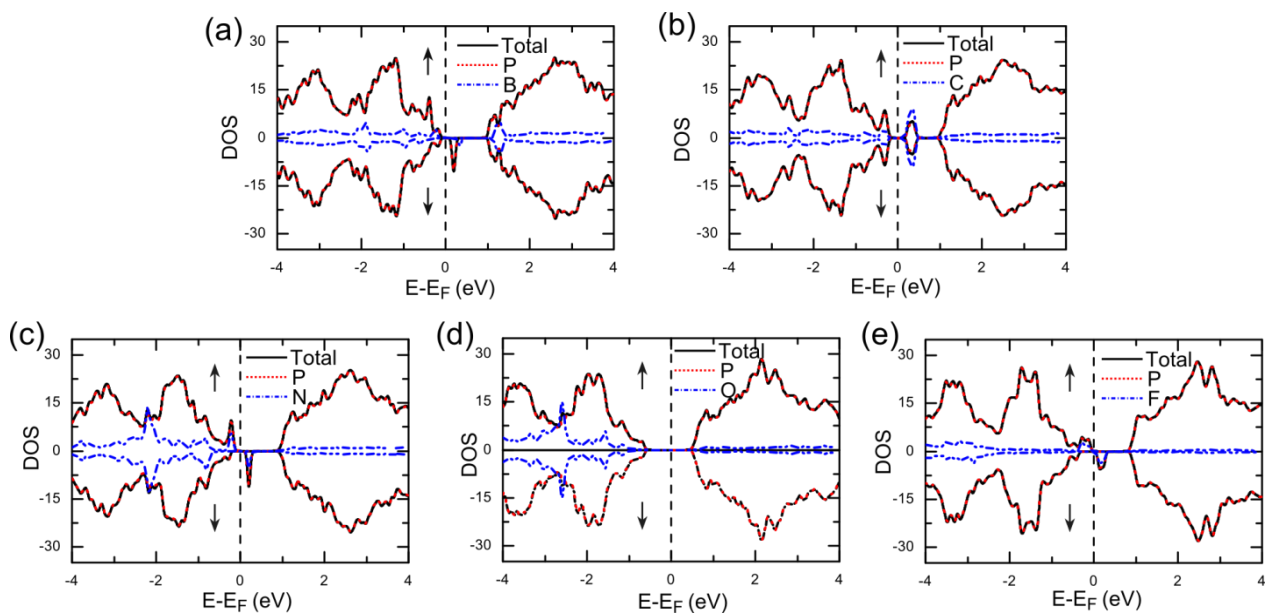

*Figure 5. Spin polarized density of states of adatoms with a [He] core on phosphorene: (a) B, (b) C, (c) N, (d) O, and (e) F. The states related to the surface adatoms are magnified by a factor of 10. The up (or down) arrow represents spin up (or spin down) density of states.*

Figure 5 displays spin and atom resolved density of states (DOS) of the adsorbed systems. B, C, N, and F will induce mid-gap states in the band gap of phosphorene. On the other hand, O will induce states inside the valence band due to the possible formation of a stronger P=O bond. Adsorption of B, N, or F also results in the spin polarized DOS inducing magnetic moment of ~1 $\mu_B$. The spin polarized charge density is found to be localized on the distorted P atoms around the adatoms (Figure S3 [34]). B has three valence electrons which form $sp^2$-like bonds with three neighboring P atoms; one of the neighboring P atoms (the one was repelled away from its original site) has an unpaired electron in the $2p$ orbit (Figure 4(a)) which induces magnetic moment in the system. The adsorption site of C is similar to that of B (Figure 4 (b)). However, C has one more valence electron than B, which could possibly pair with the electron associated with P atom resulting in the zero net magnetic moment. N and F adatoms will attract the electron from the lone pair of P leaving an unpaired electron on the $2p$ orbit of the native P atom, which contributes to the calculated magnetic moment (Figure 4 and Figure S4 [34]). O adatom will attract both of the electrons of a lone pair which will not lead to the magnetic moment in the system.

Considering that the tunneling current is sensitive to the local electronic properties of surface atoms, we now investigate the tunneling characteristics of the adatom systems. Our approach is based on BTH approximation [30], and has been successfully used to investigate tunneling characteristics of several nanomaterials including PbS quantum dot, $MoS_2$ and BN monolayers by our group [31, 41-43]. The cap of the tip used in scanning tunneling microscopy measurements (STM) was simulated by $Au_{13}$ cluster which was placed above the adatom with a distance 5 Å (Figure S5 [34]).

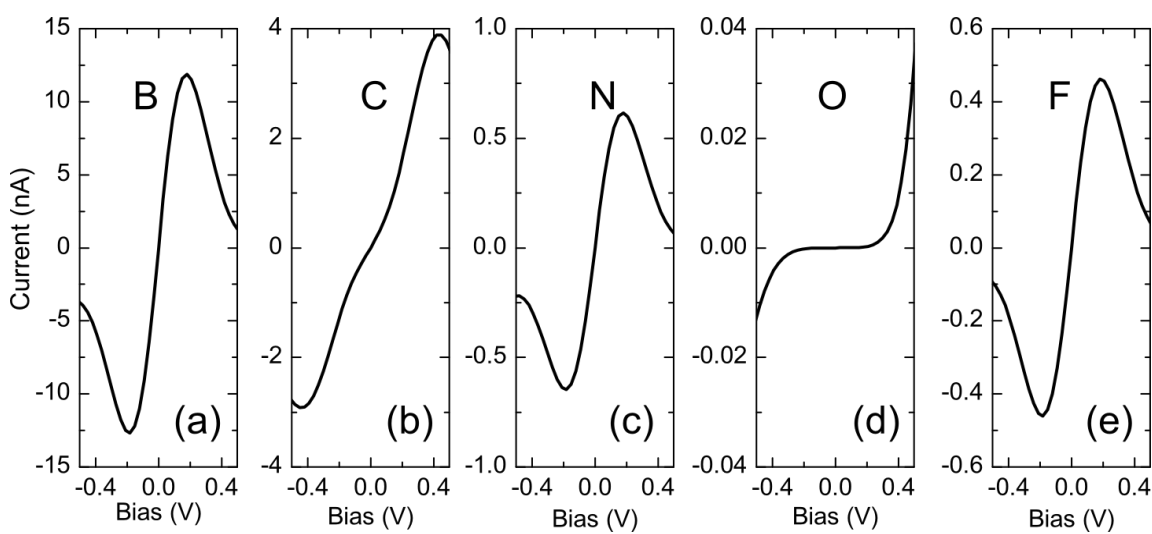

Figure 6. The tunneling characteristics of adatoms with a [He] core on phosphorene: (a) B, (b) C, (c) N, (d) O, and (e) F.

B, C, N, and F atoms show metallic tunneling characteristics with abrupt increase in the current at a small bias (Figure 6). While the tunneling current above O adatom shows a diode like behavior with a gap in the small bias region. The tunneling characteristics are consistent with the calculated DOS where mid-gap states due to B, C, N and F contribute to the tunneling current at small bias. This is not the case with O adatom since it does not introduce any mid-gap states in phosphorene. Also, prominent negative differential resistance (NDR) is observed for B, C, N and F adatoms due to the mid-gap states near Fermi level.

In summary, adsorption of light elements with a [He] core on phosphorene is investigated by using density function theory. The results find that B and C prefer the interstitial site and N, O, F atoms prefer the surface site of phosphorene. The distinct preference of these adatoms on phosphorene is the result of the interplay between electronegativity values and electronic structure of phosphorene. B, C, N, and F adsorption will induce mid-gap states leading to metallic characteristics of the phosphorene. On the other hand, oxygen adsorption is not likely to modify the electronic properties of phosphorene, and a diode like tunneling behavior is observed. Our results therefore clearly offer a possible route to tailor the electronic and magnetic properties of phosphorene by the adatom functionalization, and will be helpful to experimentalists in evaluating the performance and aging effects of phosphorene-based electronic devices.

**Acknoledgements**

The authors are grateful to Dr S. Gowtham for his support. This research was partially supported by the Army Research Office through grant number W911NF-14-2-0088. RAMA and Superior, high performance computing clusters at Michigan Technological University, were used in obtaining results presented in this paper.

*Supplementary materials for*
**Effects of extrinsic point defects in phosphorene: B, C, N, O and F Adatoms**
Gaoxue Wang, Ravindra Pandey, and Shashi P. Karna

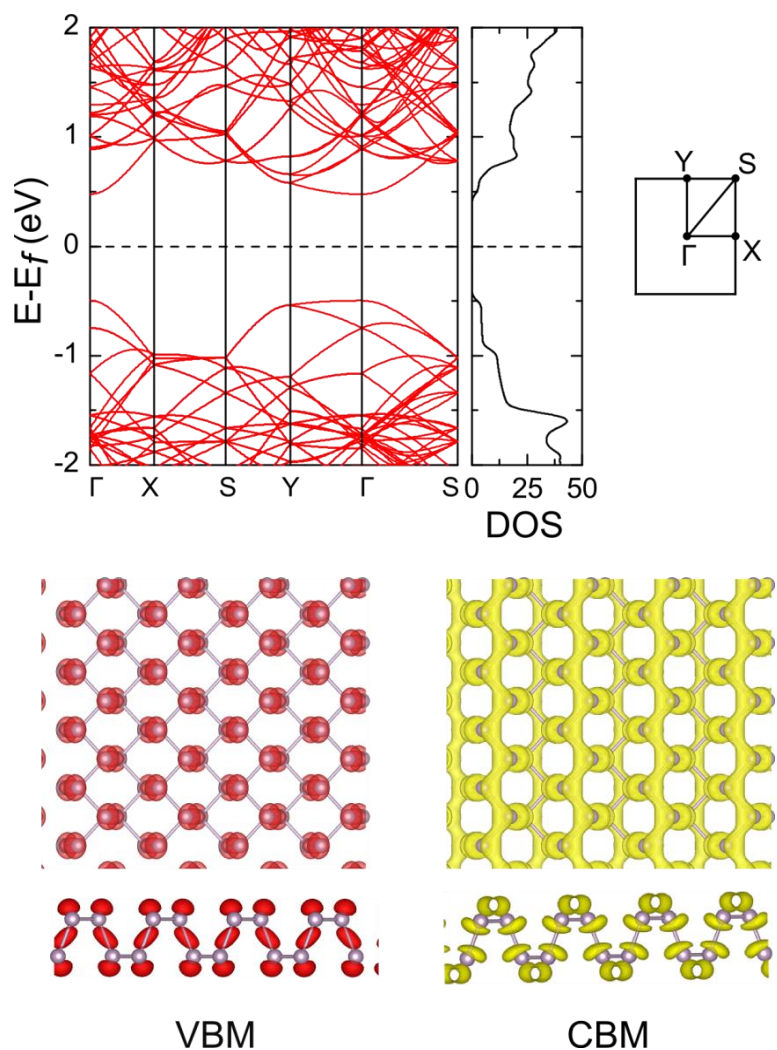

*Figure S1. The calculated band structure and density of states of pristine phosphorene. The bottom panels show the top and side view of charge density of the electronic states at valence band maximum (VBM) and conduction band minimum (CBM).*

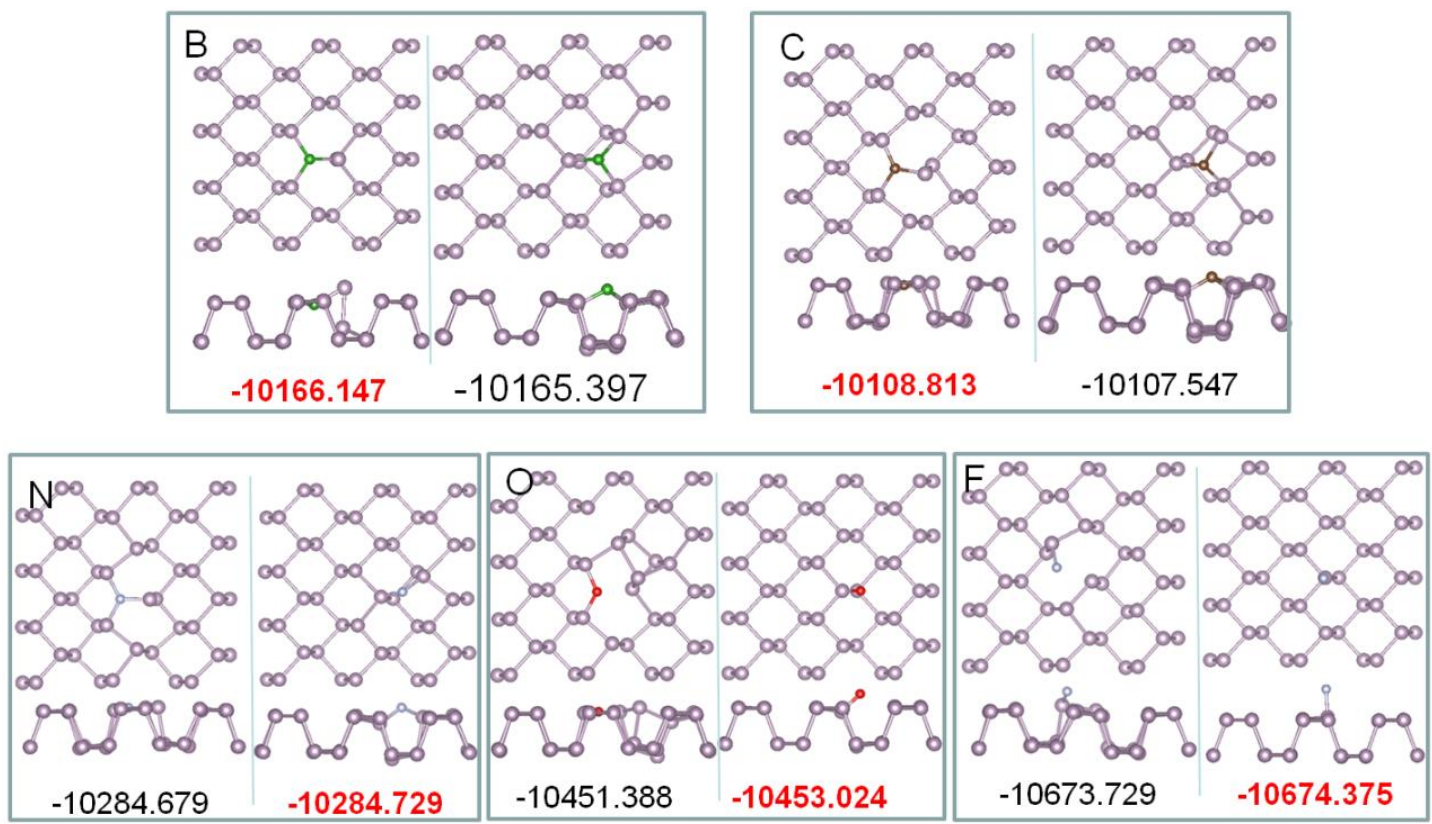

*Figure S2. The equilibrium configurations of adatoms at interstitial and the surface sites on phosphorene. The given value in eV is total energy of the configuration. B and C at the interstitial sites are more stable, while N, O and F atoms prefer to bind to the surface site.*

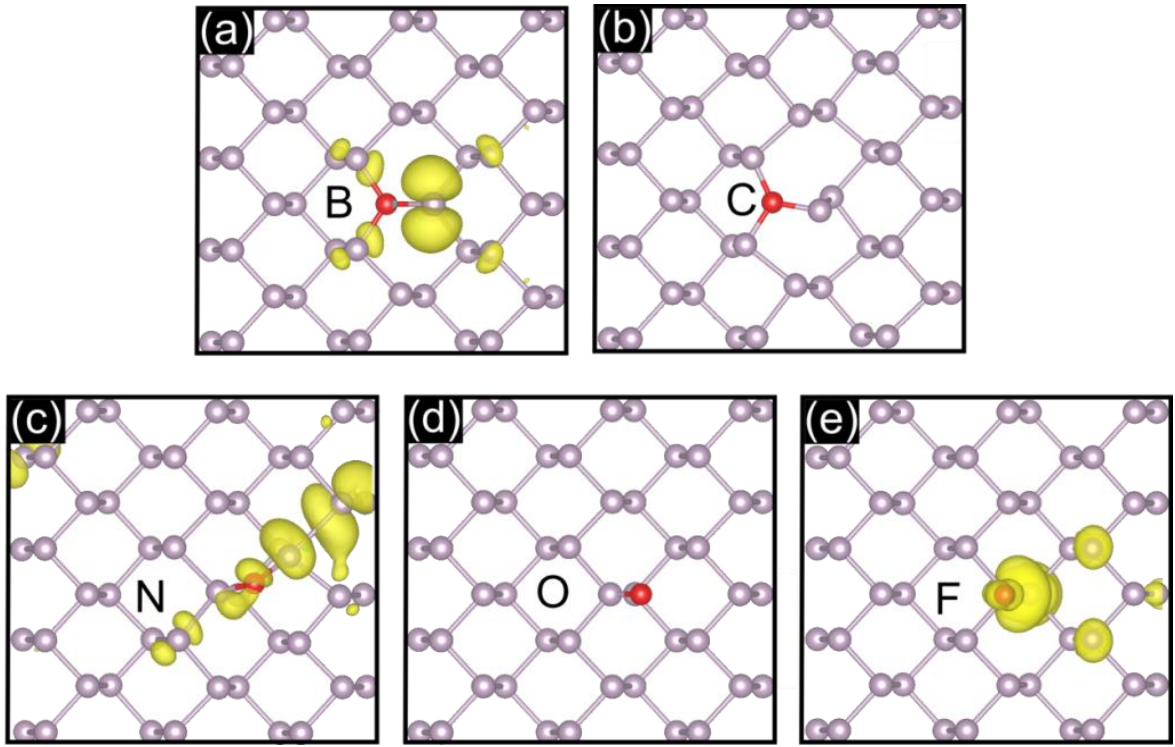

*Figure S3. Spin polarized charge density ($\rho_{up}$-$\rho_{down}$) of adatoms on phosphohrene: (a) B, (b) C, (c) N, (d) O, and (e) F. Note that C and O atoms adsorbed on phosphorene have zero magnetic moment.*

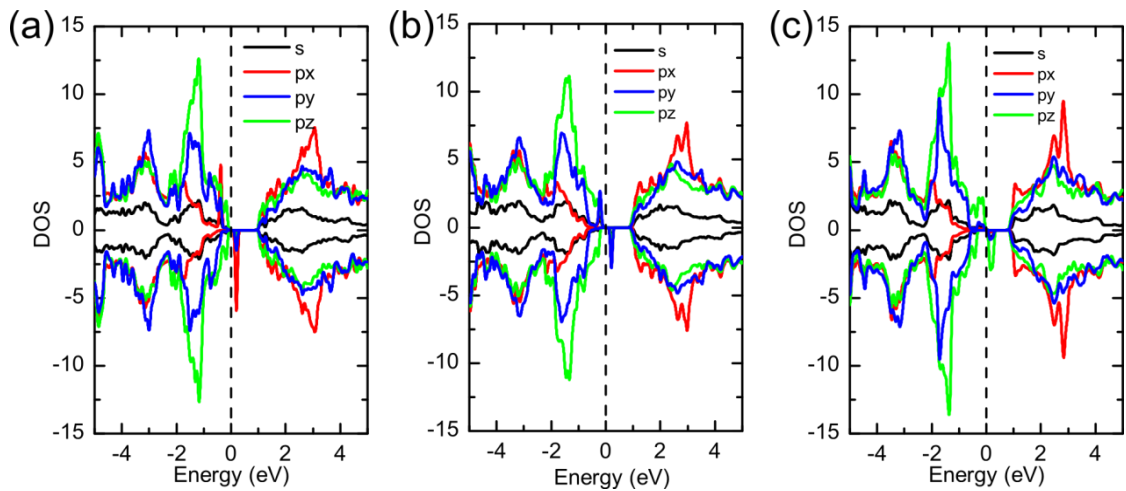

*Figure S4. Projected density of states of adatoms on phosphorene: (a) B, (b) N, and (c) F*

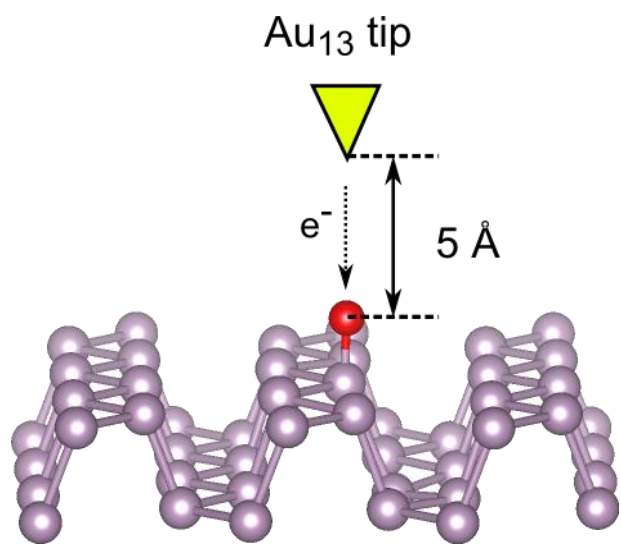

*Figure S5. Schematic illustration of the STM setup used for calculations.*

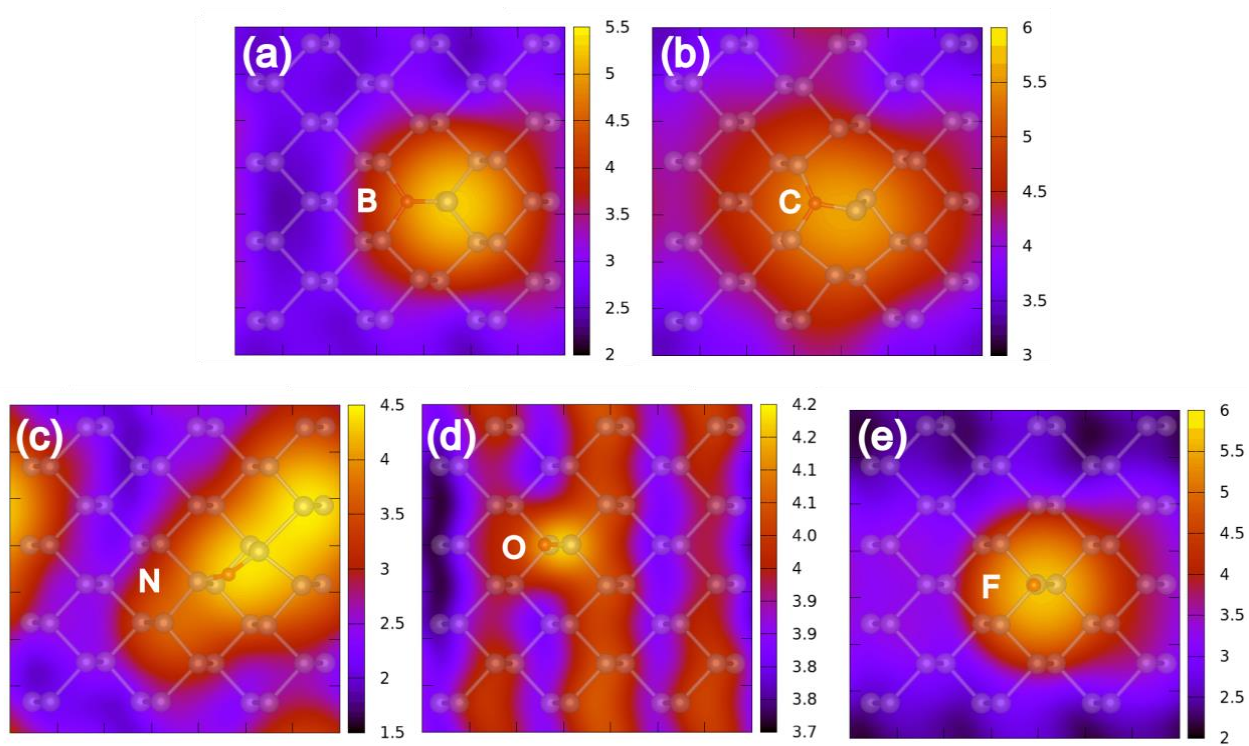

*Figure S6. Simulated STM images for adatom (B, C, N, O and F) adsorption on phosphorene. The images are simulated with a constant current of 1 nA under a positive bias of 0.5 V between the sample and the tip.*